\documentclass{aastex62}

\received{}
\revised{}
\accepted{}
\submitjournal{AJ}

\shorttitle{Ruprecht 147}
\shortauthors{Yeh et al.}

\begin{document}

\title{Ruprecht 147: A paradigm of dissolving star cluster}

\author{Fu Chi Yeh}
\affiliation{Dipartimento di Fisica Astronomia  {\it Galileo Galilei}\\
Vicolo Osservatorio 3\\
Padova, I-35122, Italy}

\author{Giovanni Carraro}
\affiliation{Dipartimento di Fisica Astronomia  {\it Galileo Galilei}\\
Vicolo Osservatorio 3\\
Padova, I-35122, Italy}

\author{Marco Montalto}
\affiliation{Dipartimento di Fisica Astronomia  {\it Galileo Galilei}\\
Vicolo Osservatorio 3\\
Padova, I-35122, Italy}

\author{Anton F. Seleznev}
\affiliation{Ural Federal University \\
620002, 19 Mira street, \\
Ekaterinburg, Russia}

\begin{abstract}
We employed recent Gaia/DR2 data to investigate the dynamical status of the nearby (300 pc), old  (2.5 Gyr) open cluster
Ruprecht~147.  We found  prominent leading and trailing tails of stars along the cluster orbit, which demonstrates that Ruprecht~147 is losing stars at 
fast pace.  Star counts indicate the cluster has a core radius of 33.3 arcmin, and a tidal radius of 137.5 arcmin. The cluster also possesses an extended corona,
which cannot be reproduced by a simple King model.
We computed the present-day cluster mass using its luminosity and mass function, and derived an estimate of  234$\pm$52 $M_{\odot}$. 
We also estimated the cluster original mass using available recipes extracted from N-body simulations obtaining a mass at birth of $\sim$ 50000$\pm$6500 $M_{\odot}$.
Therefore dynamical mass loss, mostly caused by tidal interaction with the Milky Way, 
reduced the cluster mass by about 99\%. We then conclude that Ruprecht~147 is rapidly dissolving into the general Galactic disc.
\end{abstract}


\section{Introduction} \label{sec:intro}
The vast majority of Galactic open clusters do not survive longer than a few hundredth million years (Wielen 1971; Binney \& Tremaine 1987).  Stars escape from the parent cluster because of a variety of
processes. They can be divided into internal, like two body relaxation and  stellar evolution (Lamers et al 2005; Dalessandro et al. 2015),  and external, like tidal interaction and encounters with molecular clouds (Gieles et al. 2006; Gieles \& Baumgardt 2008;
 Danilov \& Seleznev 1994).  The birth-place and initial mass also play an important role (Boutloukos \& Lamers 2003).
Signatures of dissolution can be seen in many Galactic open clusters in the form of truncated main sequences (MS) in the cluster color magnitude diagram CMD (Patat \& Carraro 1995; Piotto \& Zoccali 1999), which, in turn, produce mass functions (MF)
peaked close to high mass stars. Also, the stars distribution on the plane of the sky appears in many cases elongated or distorted.
In other stellar systems, like globular clusters or dwarf galaxies in the Local Group, one can also identify extra-tidal tail stars in the surface density profile and/or tidal tails, mainly due to the tidal interaction with
the Milky Way (Odenkirchen et al. 2003 ; Carraro et al. 2007). These features are not routinely seen in Galactic open clusters   where the low stellar density contrast makes it difficult to identify tidal escapers against the general Galactic field without performing a proper membership analysis.  
We report in this study the discovery of tidal features around the Galactic cluster Ruprecht 147.  This  
is a nearby (300 pc), old (2.5 Gyr) and solar metallicity ($[Fe/H]=0.08\pm0.07$) open cluster (Curtis et al 2013; Bragaglia et al. 2018).
Due to its proximity, accurate membership data (radial velocity and proper motion components) are available, which allow one to describe
its dynamical status  with unprecedented details.
In this work we extract astrometric and spectroscopic data from Gaia/DR2 and provide evidence of significant tidal structures around the cluster. We anticipate here that we suggest these structures are indicating the cluster suffered a conspicuous  mass loss in the past.\\

\noindent
The layout of the paper is as follows.
In Section~2 we describe data selection, and provide a list of kinematically selected members. Section~3 is dedicated to the calculation of the cluster orbit, and its relation with the tidal features. We calculate the cluster present-day mass in Section~4, while in Section~5 we provide an estimate of the cluster mass at birth. Section~6, finally, summarises our results.

\section{Data Selection} 
In order to select kinematical members of Ruprecht~147, we considered stars within 20 degrees from the cluster's center ( RA(2000.0) = 19:16:40, Dec(2000.0) = -16:17:59) having radial velocity measurements and within 500 pc from the Sun from the Gaia/DR2 archive. We imposed some conditions to ensure that all stars analysed had a good  solution from Gaia. In particular we considered stars whose relative parallax error was smaller than 20$\%$, that were observed multiple times from Gaia (visibility\_periods$>$8) and with 
 astrometric\_chi2\_al/(astrometric\_n\_good\_obs\_al-5)$<$1.44 $\times$ greatest(1,exp(-0.4$\times$(G-19.5))) as reported in Gaia Collaboration et al. (2018). This search returned 53343 stars. 
 
\begin{figure}[!t]
\centering   
\caption{Selection of proper motion and radial velocity members. Light grey histograms represent all stars, while solid histograms only stars satisfying  radial velocity constraints (upper left), and  proper motion constraints (upper right). }
\includegraphics[width=\textwidth]{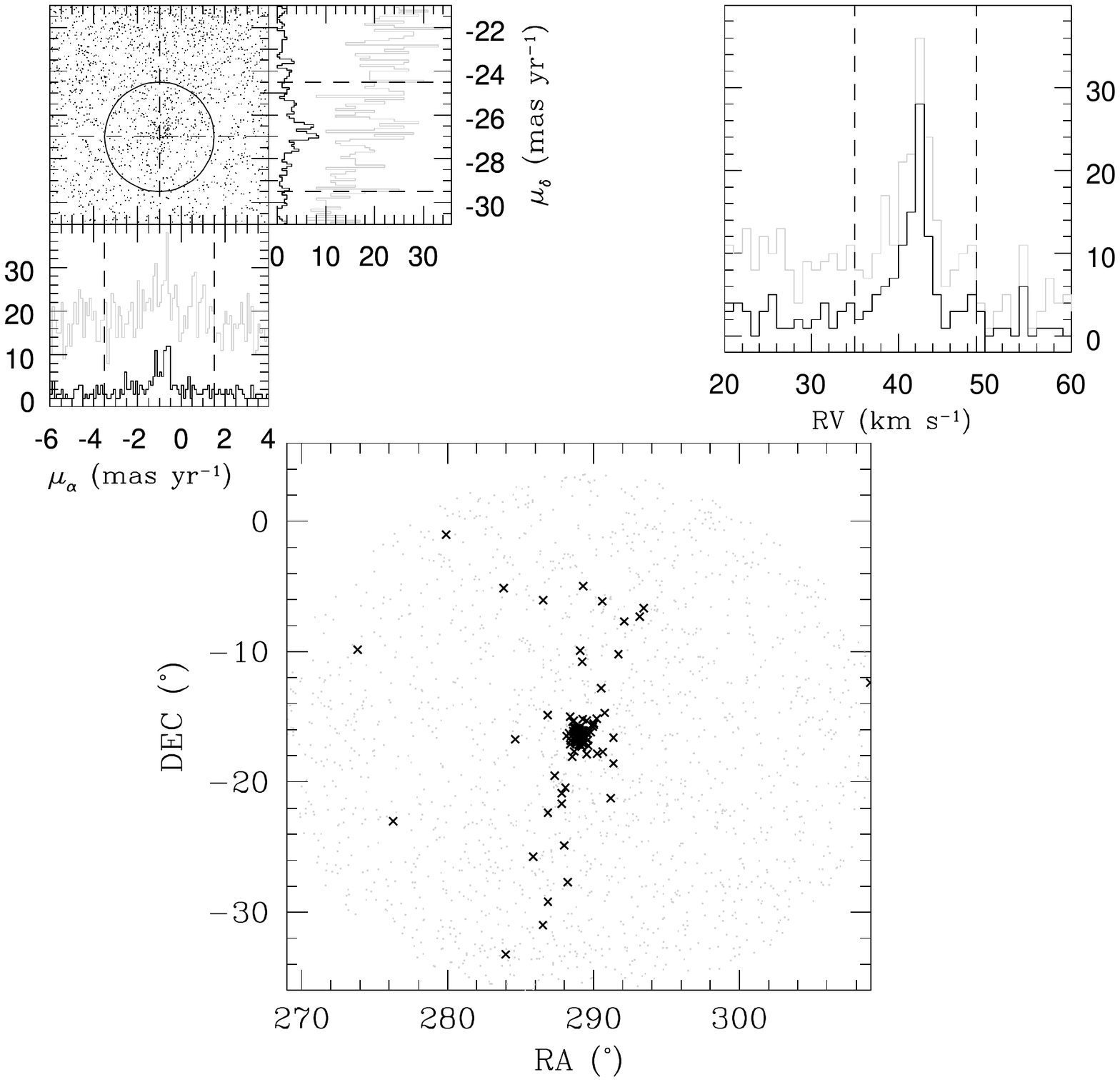}
\label{fig:PMRVEL_20DEG}
\end{figure}

We then narrowed down the selection to stars having proper motions within 2.5 mas yr$^{-1}$ from the cluster's mean proper motion ($\mu_{\alpha}=-1$ mas yr$^{-1}$
 and $\mu_{\delta}=-27$  mas yr$^{-1}$) and within 7 km s$^{-1}$ from the cluster's mean radial velocity (42 km s$^{-1}$) as shown by the dashed lines in Fig.\ref{fig:PMRVEL_20DEG}. In these plots solid lines (and symbols) denote stars that simultaneously satisfy proper motion and radial velocity selection criteria, in total
106 stars, while light gray symbols denote all other stars (proper motions where constrained between
 $-6<\mu_{\alpha} \rm [mas\,yr^{-1}]<4$
and $-21<\mu_{\delta} \rm [mas\,yr^{-1}]<-31$).

\begin{figure}[!t]
\centering   
\caption{Histograms of heliocentric Galactic coordinates and velocities of proper motions and radial velocity selected members.}
\includegraphics[width=\textwidth]{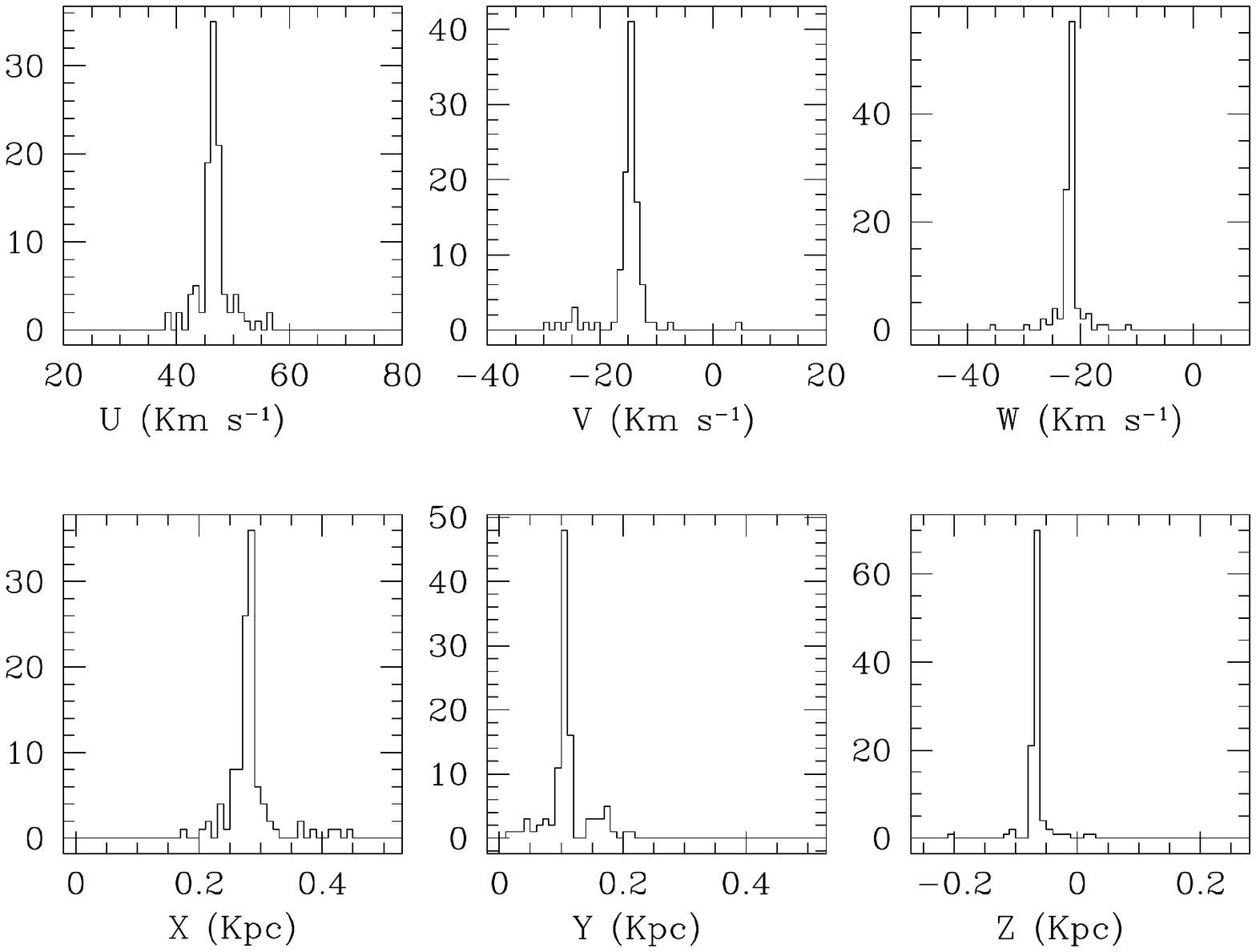}
\label{fig:HISTOGRAMS}
\end{figure}

\begin{figure}[!t]
\centering   
\caption{Selection of cluster members based on heliocentric Galactic velocities, colours and magnitudes. Crosses indicate candidate members satisfying kinematic selection criteria and with photometry consistent with cluster membership. Magenta crosses indicate in particular candidate members beyond 3 degrees from the cluster center. Red open squares or black open circles are candidate non-members based on their photometry (circles) or kinematics (squares).}
\includegraphics[width=\textwidth]{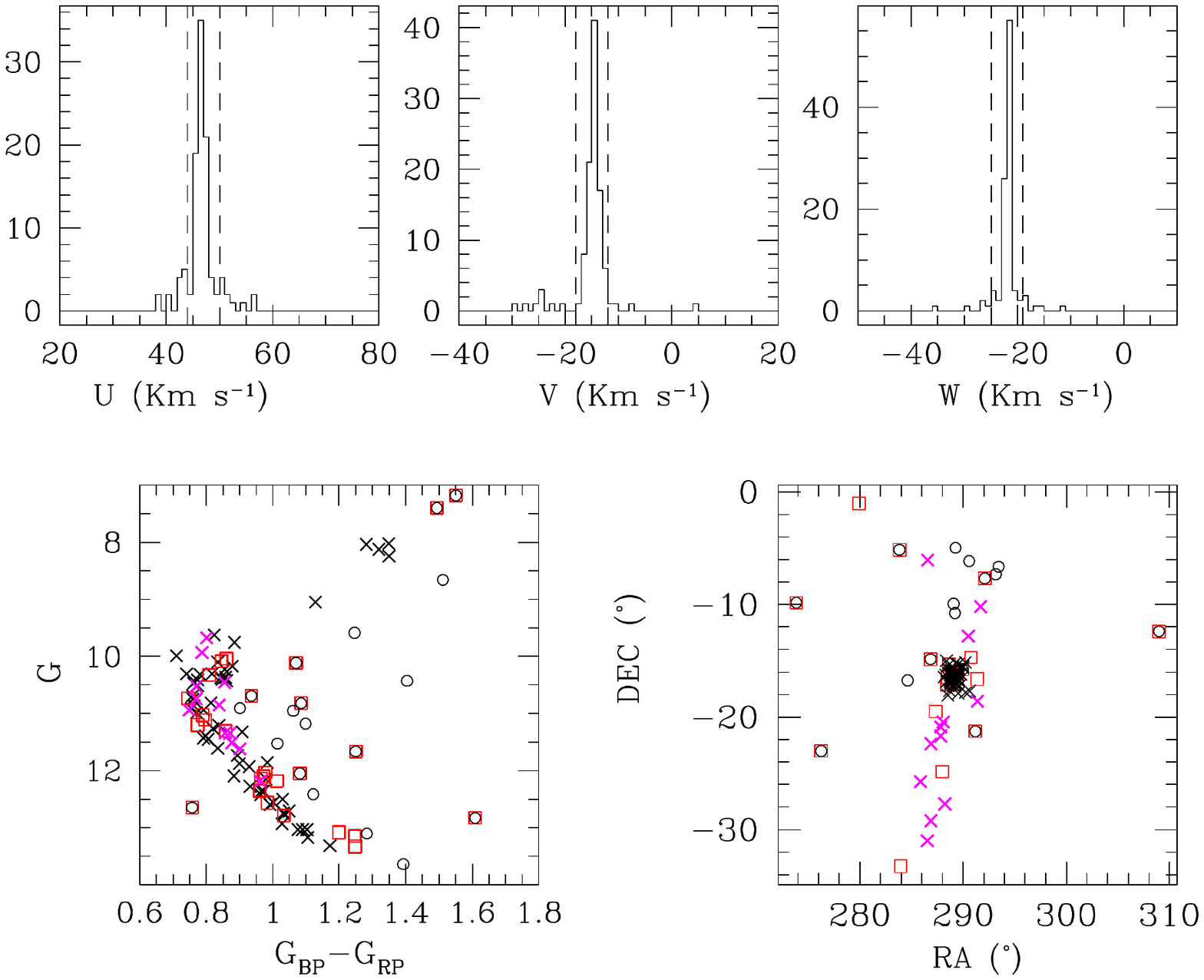}
\label{fig:SEL}
\end{figure}

\begin{figure}[!t]
\centering   
\caption{Different tridimensional projections of the selected members.}
\includegraphics[width=0.7\textwidth]{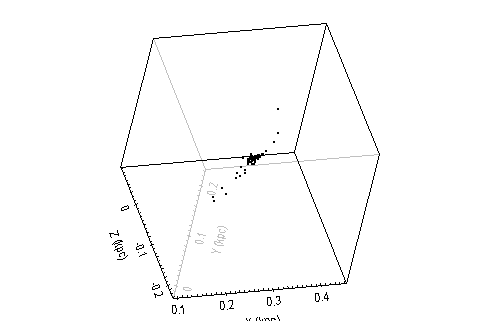}
\includegraphics[width=0.7\textwidth]{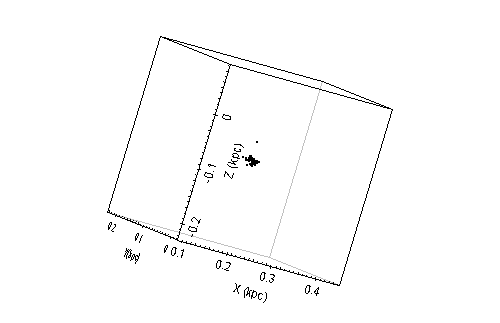}
\includegraphics[width=0.7\textwidth]{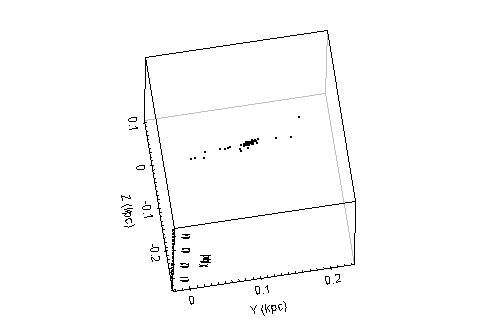}
\label{fig:XYZ}
\end{figure}

We calculated then the heliocentric Galactic velocities and further 
restricted the sample to stars having velocity within 3 km s$^{-1}$
from the mean cluster's velocity (see Table 1 and Fig.\ref{fig:HISTOGRAMS}) as denoted by the dashed lines in Fig.\ref{fig:SEL}. Stars that do not satisfy this kinematic criterion are indicated by the red open squares in the bottom panels of Fig.~\ref{fig:SEL}.
We also clipped stars whose colours and magnitudes appeared inconsistent with cluster's membership (as shown by the open circles in the bottom panels of Fig.\ref{fig:SEL}). Stars which survived these selection criteria are indicated by the crosses
in the same Figure, where magenta crosses indicate stars beyond 3 degrees from the cluster's center.  After these selections we obtained a sample of 69 candidate members. Fig.\ref{fig:XYZ} shows different tridimensional spatial projections of the selected members. 

\begin{table}[!ht]
	\centering
	\begin{tabular}{ c c c c c c} 
	\hline
     X & Y & Z & U & V & W\\
     (kpc) & (kpc) & (kpc) & (km s$^{-1}$) & (km s$^{-1}$) & (km s$^{-1}$) \\
     \hline
    0.28$\pm$0.04 & 0.11$\pm$0.04 & -0.07$\pm$0.02 & 47$\pm$3 & -15$\pm$4 & -22$\pm$2 \\
	\end{tabular}
    \caption{Mean heliocentric Galactic coordinates and velocities}
    \label{tab:table}
\end{table}

\noindent
A close inspection of Fig.\ref{fig:SEL} (lower right panel) indicates that the cluster is made of a central concentration of stars, 
roughly elongated in the Declination direction,  and of two sparse, similarly elongated,  structures, which resembles a leading and a trailing tidal tail. These two tails (showed by magenta crosses) 
are mostly made by MS stars, as one can appreciate looking at the cluster CMD in the lower left panel. These are therefore photometric and astrometric cluster members which beyond any reasonable doubt are leaving the cluster.

\section{Cluster orbit}
\begin{figure}[!t]
\centering   
\caption{Different tridimensional projections of the selected members. The solid red line is the cluster actual orbit. Axis are in kpc.}
\includegraphics[width=0.9\textwidth]{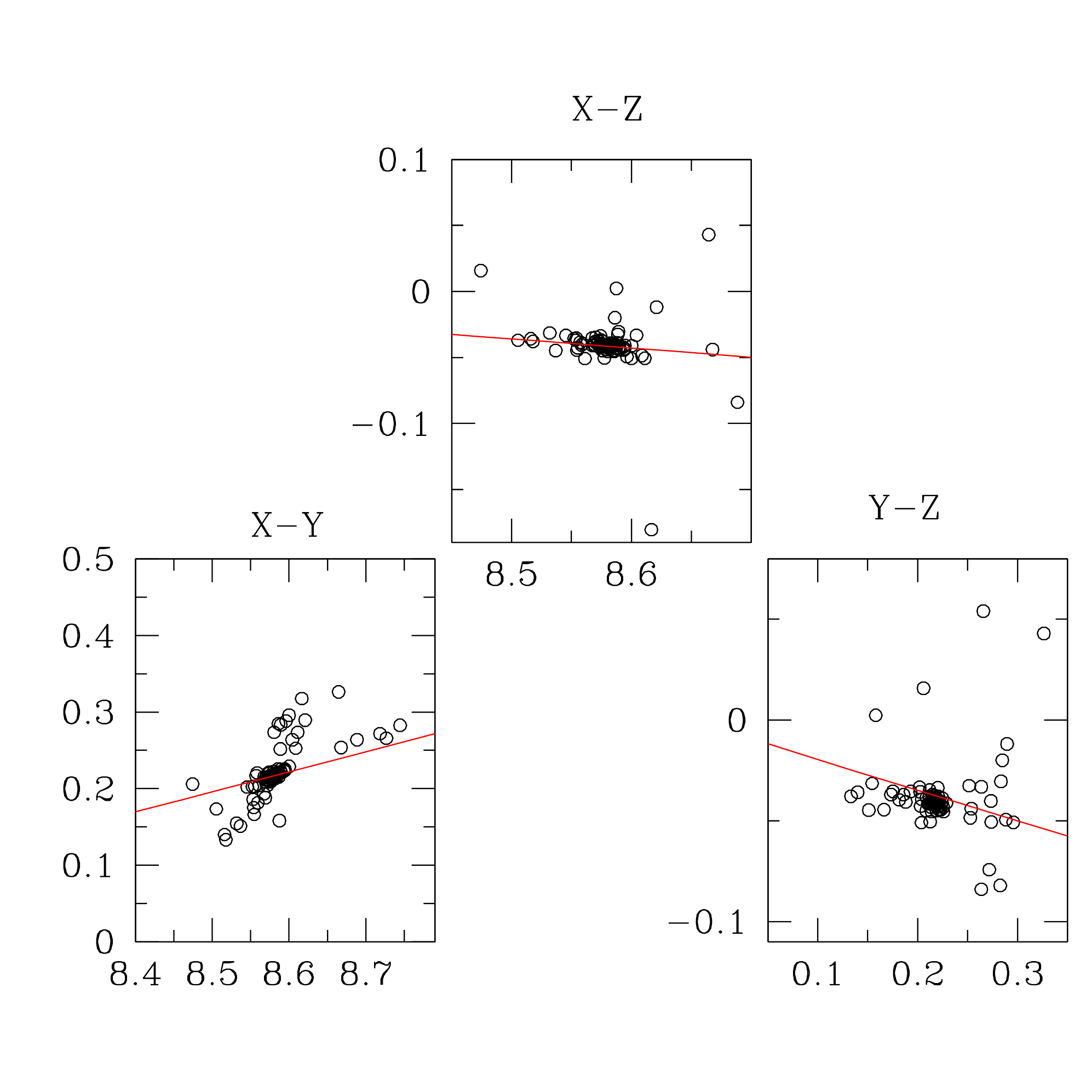}
\label{fig:orbit}
\end{figure}

A further confirmation of the tidal origin of these structures  would be to show that  they are located along the cluster present-day orbit.
To this purpose, we take from 
Table~1  the Cartesian velocity components U , V , and W, which we shifted  to the local standard of rest and corrected
for solar motion. We then   compute Ruprecht~147 orbit using test particle simulations in a fixed potential. This is justified by the fact that the integration is short in time and essentially aimed at deriving the present day orbit and orbital parameters.
The details of the orbit integration, and original code parameters are in Carraro et al. (2002).
Briefly, to integrate the orbit of Ruprecht~147 we adopted  a modified Allen \& Santillan (1991) model for the Milky Way (MW) gravitational potential. This potential is time-independent, axisymmetric, fully analytic, and mathematically very simple. 
We normalised the assumed densities for the bulge, disk, and halo so that the combined gravitational force fits a rotation curve consistent with most recent observations,
and Galacto-centric distance and rotation velocity for the Sun (Bland-Hawthorn \& Gerhard 2016, Reid et al. 2014). 
For this simple exercise, we did not include any time-dependent component, like bar or spiral arms, which we believe are not  very important for an object located so close to the Sun.
Besides, it is reasonable to believe that the Galactic potential did not change much over about a Galactic rotation time ($\sim$ 250 Myr), which is more than enough for the purpose of deriving Ruprecht 147 present day orbit.
The orbit-integration routine consists of  a fifteenth-order symmetric, symplectic Runge-Kutta method, using the Radau scheme (Everhart 1985). This guarantees conservation of energy and momentum at a level of $10^{-12}$ and $10^{-9}$, respectively, over the whole orbit integration. The orbit, integrated back in time for 200 Myr, is  shown in Fig~\ref{fig:orbit}.
This figure clearly shows that the tidal features associated to Ruprecht~147 distribute along the cluster orbit. The 
projection where tails are closer to the cluster's orbit seems to be the Y-Z. This confirms recent results obtained by Roser et al. (2018) and Meingast \& Alves (2018) for the Hyades star cluster.

\section{Present-day mass estimate}
The alignment of tidal tails  along the cluster orbit indicates that Ruprecht~147 is losing mass due to tidal interaction with the Milky Way gravitational potential. 
Additional indications of mass loss should be hardcoded in the cluster mass function, since it is very well-known that clusters in advanced stages of dynamical evolution routinely possess MSs
which appear depleted of low mass stars (see, e.g., Patat \& Carraro 1995; Piotto \& Zoccali 1999).
In order to obtain a reliable estimate of the cluster present-day mass we relaxed the magnitude and radial velocity constraints and
performed star counts using stars brighter than G =18 mag within a larger field (40 degrees from the cluster centre)  and having proper motion components within 2.5 mas from the means (see Section~2),
and, finally, parallaxes between 2 and 4.5.  
This way we obtained a  sample of 3739 stars. \\

\begin{figure}[!t]
\centering   
\caption{Surface density map in a region of 40 degrees on a side around Ruprecht~147.}
\includegraphics[width=0.9\textwidth]{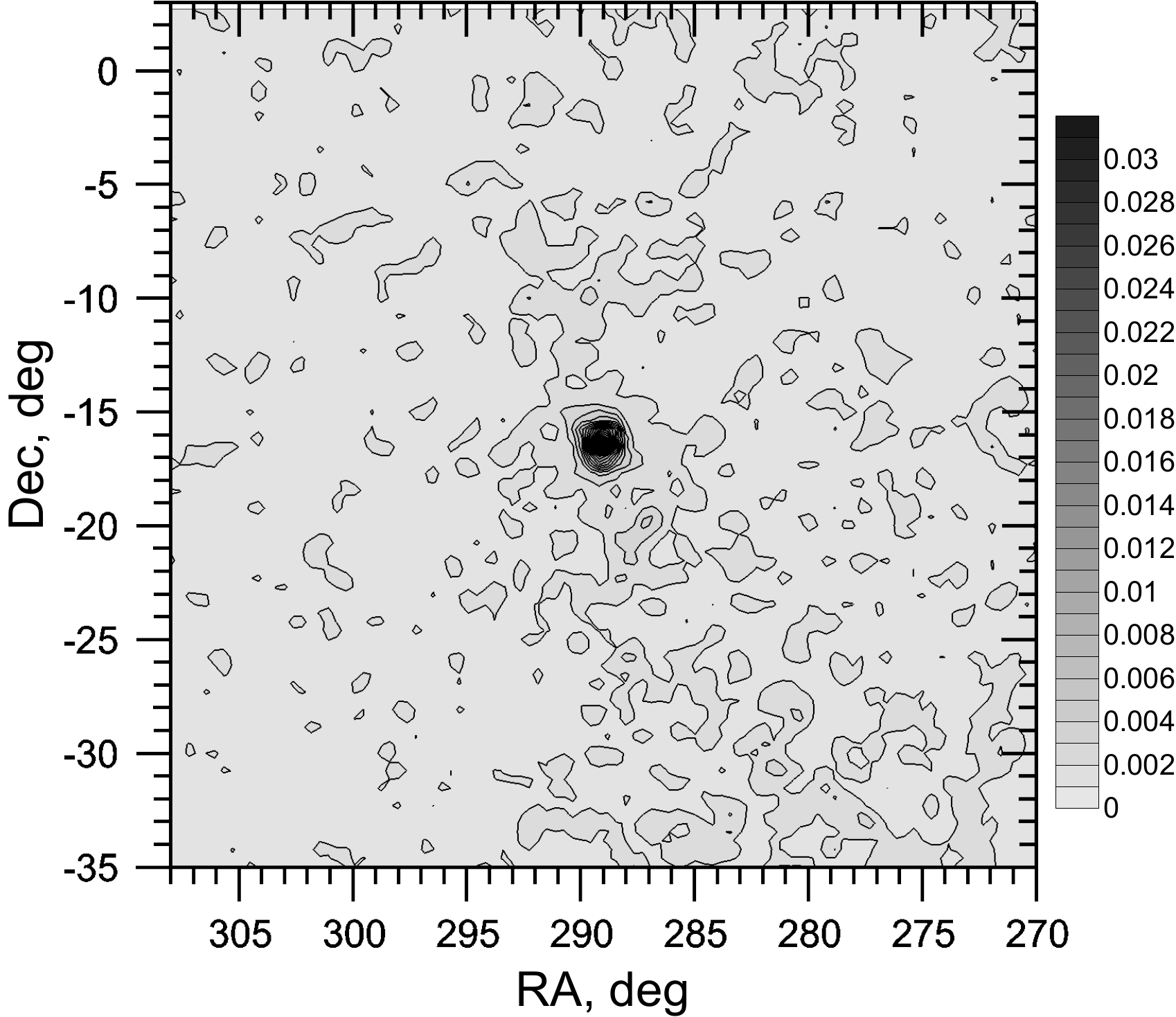}
\label{fig:sdm}
\end{figure}

\noindent
From this sample we computed a  surface density map using a kernel density estimator (KDE) technique with a  kernel
density halfwidth of 40 arcmin following the method detailed  in 
Carraro \& Seleznev (2012) and in Seleznev (2016b). 
The resulting density map is shown in Fig.~\ref{fig:sdm}.
By closely inspecting this map one can notice that the densest part of the cluster is slightly elongated in the
north-south direction thus indicating some tidal deformation. Also, the
stellar density outside the densest part of the cluster reveals the same elongated structure which coincides with the cluster tidal
tails (see previous Sections).

\begin{figure}[!t]
\caption{Linear (left panel) and logarithmic (right panel) radial surface  density profile  in a region of 20 degrees  around Ruprecht~147.}
\includegraphics[width=0.5\textwidth]{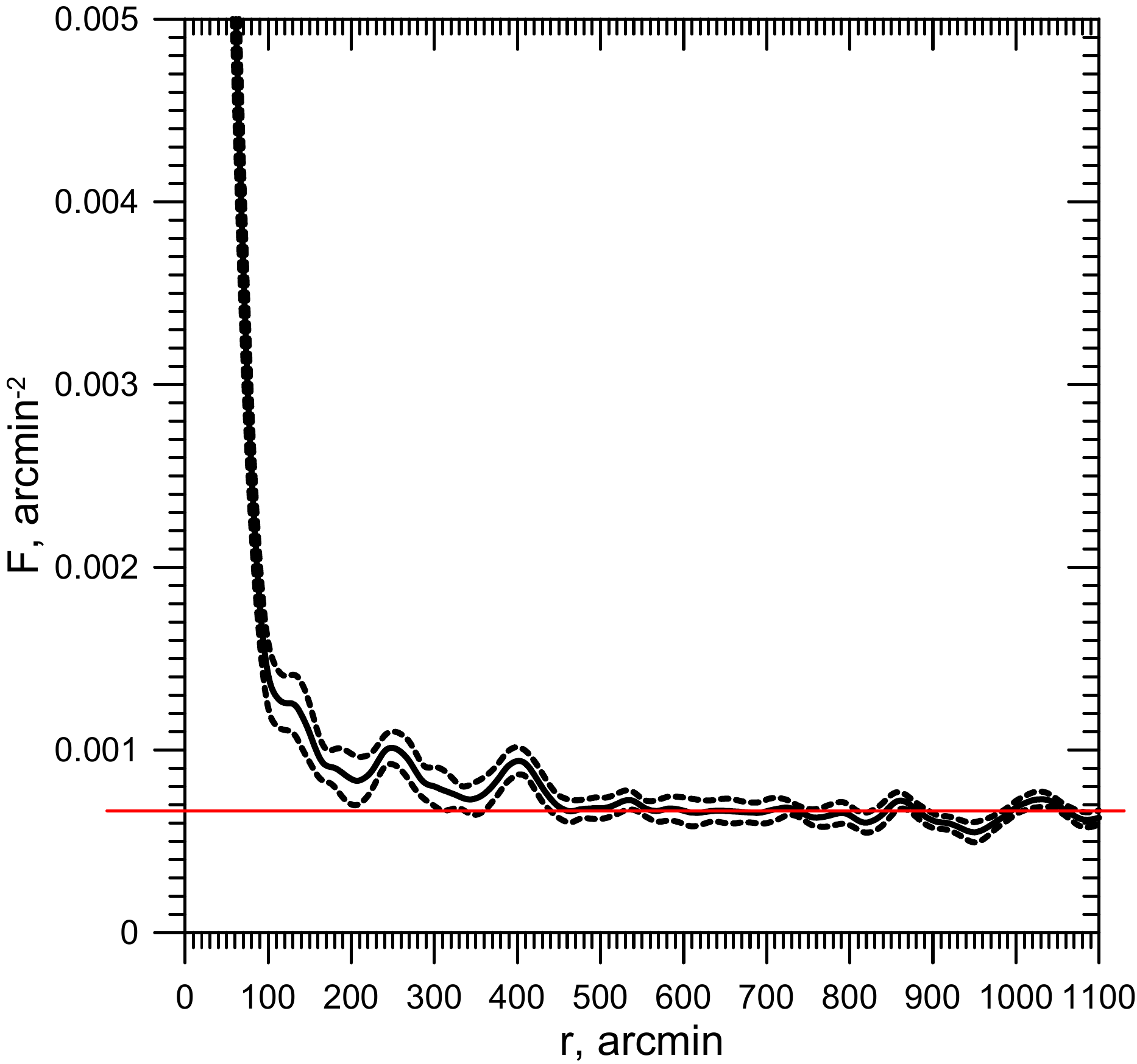}
\includegraphics[width=0.5\textwidth]{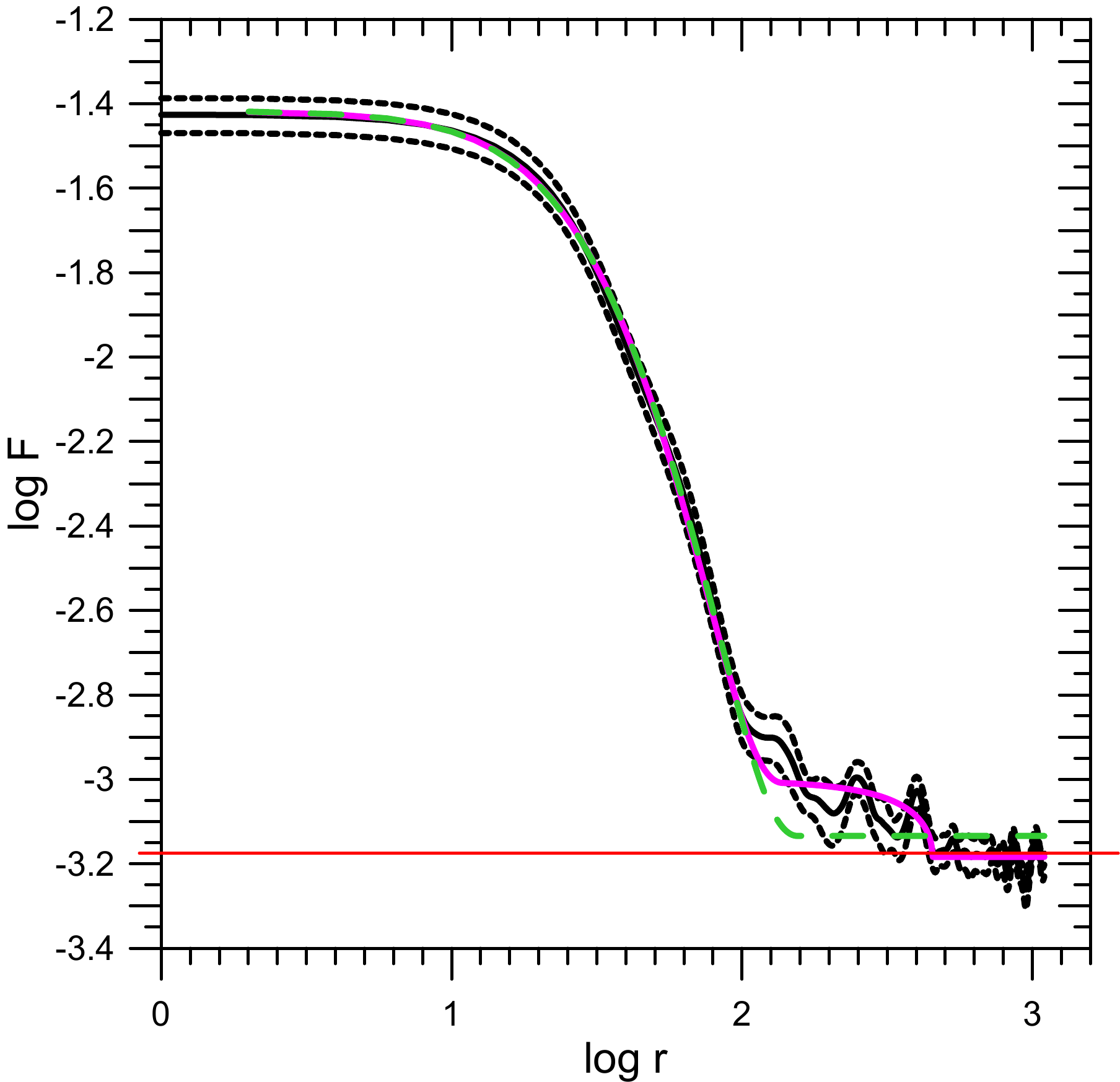}
\label{fig:sdp}
\end{figure}

The corresponding linear and logarithmic  radial density profile  are
shown in Fig.~\ref{fig:sdp}, where the red solid line indicates the level of the background.
On the other hand, dashed lines indicate the 2-$\sigma$ uncertainty  in star counts.
The cluster structure shows a well-defined
King-like cluster central concentration (see right panel) having a radius of about 100 arcmin,  and an  extended
cluster corona, up to roughly 450 arcmin from the centre (see also the left panel). The corona consists of stars that left the cluster but are still
moving in its vicinity (Danilov et al. 2014).
Following the approach described in Seleznev (2016a) we can estimate the
mean density of the field stars as $F_b=0.0007\pm0.0001 \; arcmin^{-2}$ and the
cluster corona radius as $R_c=450\pm10 \; arcmin$.\\

In the right panel of Fig.~\ref{fig:sdp} we perform a fit with a King (1962) model (green dashed line)  and with a modified King model (Seleznev 2016a) which takes into account the presence of a corona
around the cluster (magenta solid line).
Clearly, the simple King model does not reproduce the cluster corona. In the case of the modified King model 
we found that the cluster has a core radius $r_c = 33.3\pm0.2$ armin and a tidal radius $r_t = 137.5\pm1.7 armin$. 
With these estimate the King (1962) concentration parameter would be c = 0.62.\\

\noindent
The integration of the density profile yields an  estimate of the cluster star
number of $N_{cluster}=280\pm100$, while the star number within 100 arcmin is $N_{core}=160\pm20$.\\

\begin{figure}[!t]
\centering   
\caption{Luminosity functions of Ruprecht 147 and the surrounding stellar field.}
\includegraphics[width=0.5\textwidth]{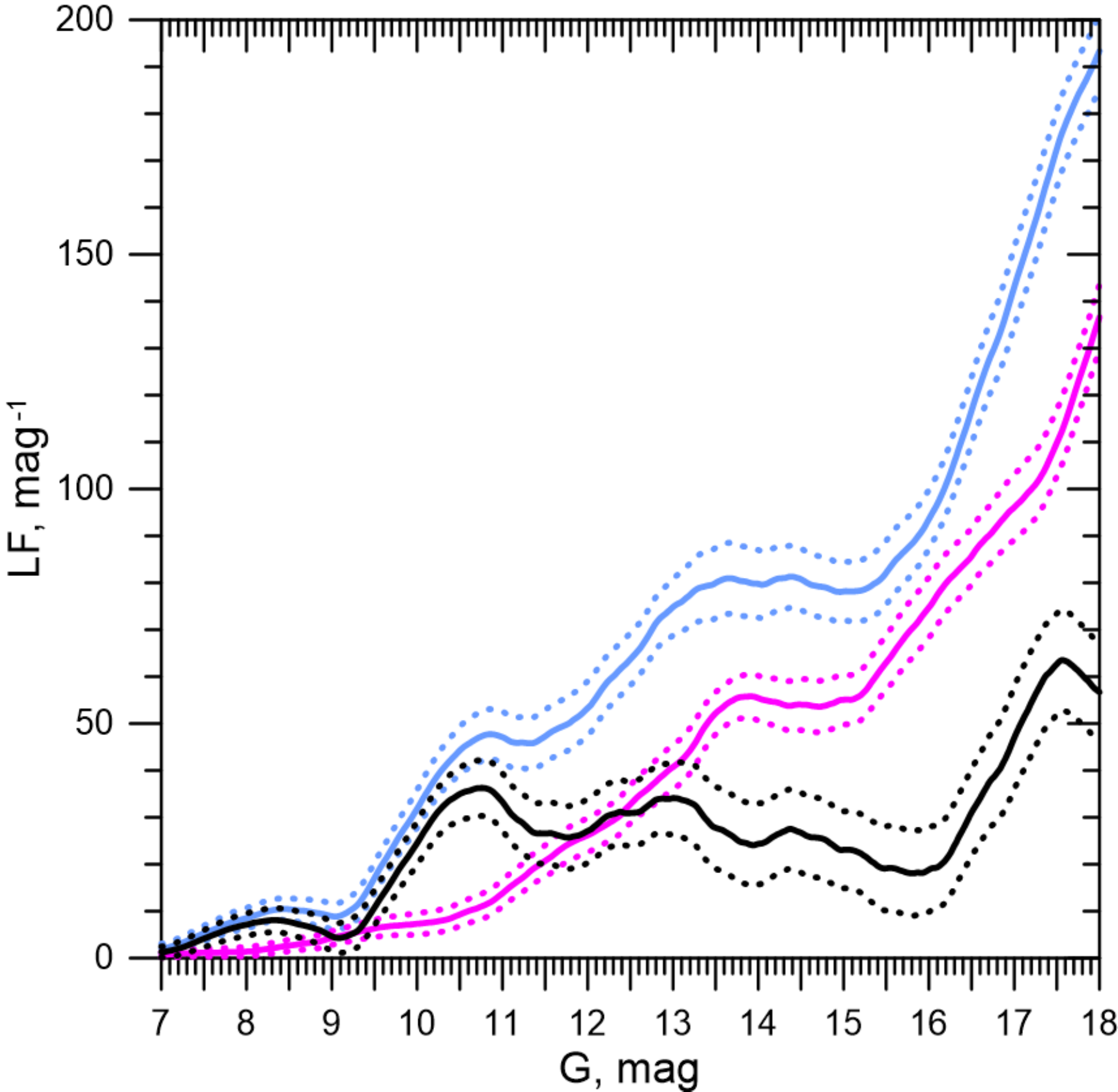}
\label{fig:lf}
\end{figure}

\noindent
The luminosity function (LF) of the cluster was then estimated following 
Seleznev (1998) and Seleznev et al. (2000). A ring around the cluster with the same area as the cluster
was taken as reference field. The LF is shown in Fig.\ref{fig:lf}.  The solid black line
shows the field star subtracted cluster LF.
Dashed lines, as before, indicate 2-$\sigma$ uncertainty in star counts.
For a better understanding we also show the whole cluster LF (in blue), and the equal area comparison field LF (in purple). \\

\begin{figure}[!t]
\centering   
\caption{Linear (left panel) and logarithmic (right panel) mass function of Ruprecht~147}
\includegraphics[width=0.4\textwidth]{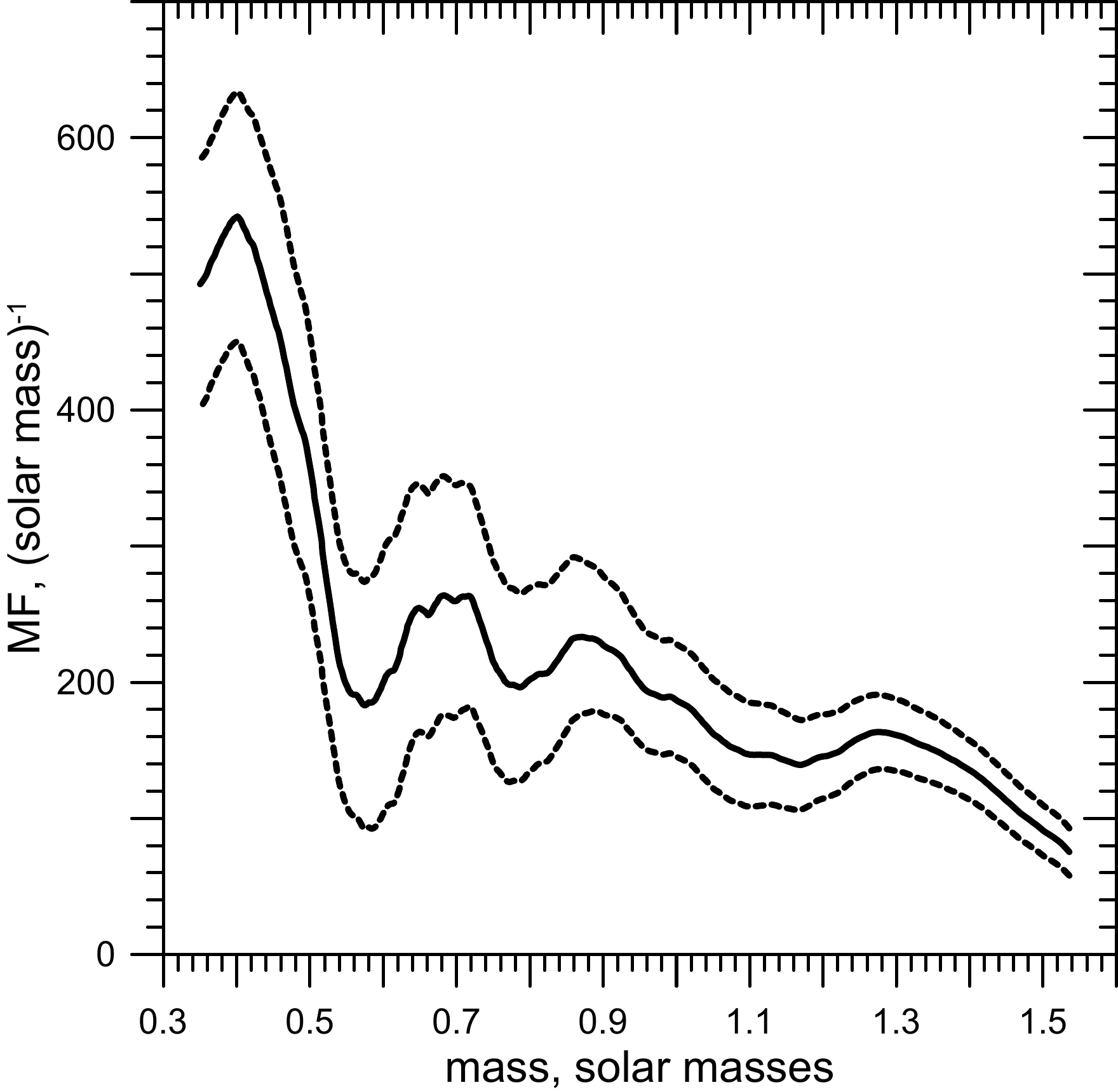}
\includegraphics[width=0.4\textwidth]{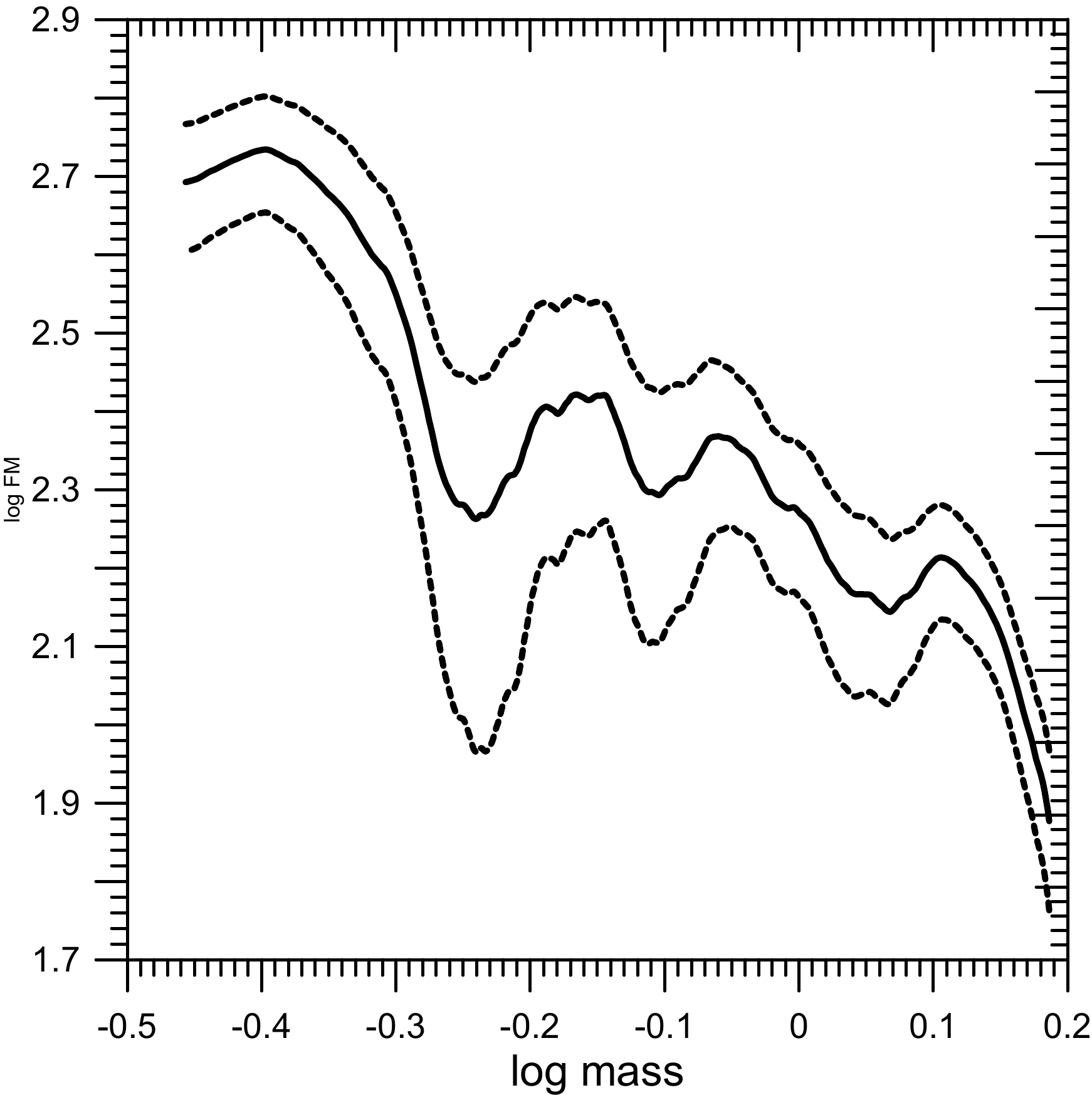}
\label{fig:mf}
\end{figure}

\begin{figure}[!t]
\centering   
\caption{Mass functions of Ruprecht 147. See text for details.}
\includegraphics[width=0.7\textwidth]{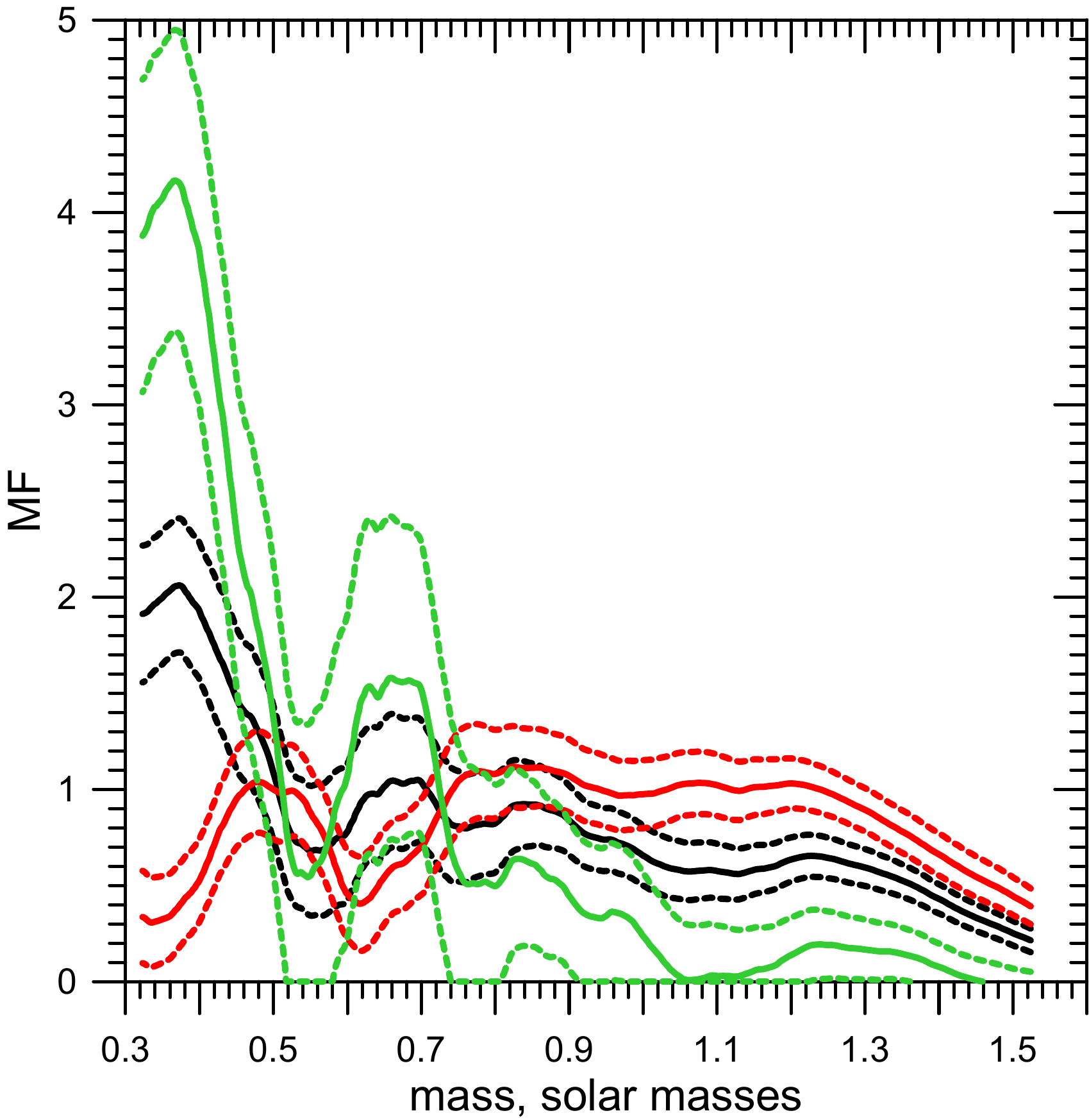}
\label{fig:mffin}
\end{figure}

\noindent
The  corresponding cluster mass function (MF) was then computed following 
Seleznev et al (2017). The linear and logarithmic 
MFs are shown in Fig.\ref{fig:mf}. The mass range corresponds
to stars with $G\in[9.76;18.0]$ magnitudes, and the bright evolved cluster stars are
excluded due to the complicated mass-luminosity relation for this kind of stars.  As usual, dashed lines
indicate 2-$\sigma$ uncertainty. 
Finally, the MFs
of the whole cluster (in black), the cluster core (in red) , and the cluster corona (in green) normalised to unity are
then shown in Fig.\ref{fig:mffin}. This figure
clearly highlights mass segregation, namely  the central region has a deficiency in the low-mass
stars compared to the cluster and its corona.\\

\noindent
The cluster present-day mass was finally inferred by integrating the MF. The number of
stars with $G\in[7.0;9.76]$ magnitudes was estimated via the cluster and the cluster
core LF integration in this magnitude range. They amount to be $18\pm7$ for the whole
cluster (the cluster core and the cluster corona) and $19\pm6$ for the cluster core
(it basically means that all bright stars are inside the cluster core). The mean mass of these
stars can be estimated as 1.55 solar masses by theoretical isochrone table (Bressan et al. 2012) for
the age of 2.5$\pm$0.5 Gyr (Curtis et al. 2013; Bragaglia et al 2018). Finally, we obtained estimates of the cluster star number $N_{cluster}=280\pm66$
and the cluster mass $M_{cluster}=234\pm52 M_{\odot}$  (the cluster core and corona).
The same estimates for the cluster central region are $N_{c}=159\pm32$,  and $M_{c}=163\pm28 M_{\odot}$.
These anyway have to be considered as lower limits, since we did not take into account stars fainter than G = 18, invisible remnants of evolved massive stars,
and unresolved binary or multiple stars.\\
The mass we would obtain by considering only cluster members selected as in Section~2 is about 110 $M_{\odot}$.

\section{Initial mass estimate}
The observational evidences presented in the previous sections reveal that Ruprecht~147 might have lost a significant fraction of its original mass due to environmental effects (like tidal shocks due to close interactions with giant molecular clouds, spiral arms, the Galactic disc and, in general, to the interactions with the Galactic tidal field) or internal dynamical effects (like two-body relaxation or stellar evolution). In this section, we attempt to estimate the total mass lost by Ruprecht ~147 during its evolution and derive its original mass.
We adopted here the approach described by Lamers et al. (2005), and already used by Dalessandro et al.(2015) to derive the initial mass of the star cluster NGC~6791. This method  has the advantage of describing the way the mass of a cluster decreases with time by means of relatively simple analytic expressions, extracted from a large suite of N-body models.
The initial mass can be estimated by the expression (see Dalessandro et al. 2015):

\begin{equation}
M_{ini} \sim [(\frac{M}{M_{\odot}})^{\gamma} + \frac{\gamma t}{t_o}]^{\frac{1}{\gamma}} [1-q_{ev}(t)]^{-1}
\end{equation}

\noindent
where t is the cluster age and M its present-day mass. The function $q_{ev}(t)$ describes the mass loss produced by stellar evolution, and essentially depends on metallicity. For a mildly super-solar   star cluster ([Fe/H] = 0.08$\pm$0.07, see Bragaglia et al. 2018), from Lamers et al. (2005) we read: 

\begin{equation}
log_{10} q_{ev} (t) = (log_{10} t - 7.0)^{0.255} -1.805.
\end{equation}

\noindent
Finally, the index $\gamma$ and $t_o$ depend on model cluster initial profiles and the tidal field, which are taken to be King-like (King 1962). 
Following again Dalessandro et al. (2015) and Lamers et al. (2005) we adopt  here $\gamma$=0.62 and t$_o$ = 3.3 Myr.  In fact, as described in Dalessandro et al. (2015),  t$_o$ is a constant
depending on the strength of the tidal field. In other words, it does not depend on a particular cluster model. Since Ruprecht~147 lies at the same Galacto-centric distance of both the Sun and NGC~6791, this seems a reasonable assumption. As for $\gamma$, we adopted 0.62 because this is a typical value for Galactic open clusters (Baumgardt \& Makino 2003). 

\noindent
By inserting all the numerical values we obtain from Equation~1 an estimate of the cluster initial mass of M $\sim$ 50000$\pm$6500 M$_{\odot}$, where the reported uncertainty has been derived through propagation of age and actual mass errors.
This would imply that the cluster lost about {\bf 99\%} of its original mass over its life.

\section{Conclusions}
In this study we made use of literature data to investigate the dynamical state of the nearby, old open cluster Ruprecht~147.
We started from the evidence that the  spatial distribution of star cluster members indicate the presence of prominent tidal features. We found that these features are aligned with the cluster orbit, which we calculated using a simple Galactic model. Due to the short integration time this is entirely reasonable.\\
\noindent
From a sample of 3739 proper motion and parallax selected members we obtained an estimate of the actual cluster mass of $234\pm52 M_{\odot}$, and compared it with an estimate of the cluster initial mass of $\sim 50000\pm6500 M_{\odot}$, derived from analytical interpolations of N-Body simulation results. 
The evidence emerges that Ruprecht~47 lost more than {\bf 99\%} of its original mass over its lifetime.  
Given the presence of significant tidal tails we suggest that most of this mass loss is of tidal origin,  and that the cluster is undergoing fast dissolution into the general Galactic disc.

\acknowledgments
Fu Chi Yeh acknowledges the European Union founded Astromundus program ({\tt https://www.uibk.ac.at/astromundus/}), under which she could  spend the 2018 spring semester in Padova.
The comments of an anonymous referee are deeply acknowledged. The work of A.~F. Seleznev was supported by the Ministery of Science and Higher Education (the basic part of the State assigment,
RK no AAAA-A17-117030310283-7) and by the act no 211 of the Government of the Russian Federation, agreement no. 02.A03.21.0006 .

\vspace{5mm}
\facilities{Gaia}

\software{}

\end{document}